# Gigantic Anisotropy of Self-Induced Spin-Orbit Torque in Weyl Ferromagnet Co$_2$MnGa

Motomi Aoki[1,2], Yuefeng Yin[3,4], Simon Granville[5,6], Yao Zhang[5,6], Nikhil V. Medhekar[3,4], Livio Leiva[1], Ryo Ohshima[1,2], Yuichiro Ando[1,2,7], and Masashi Shiraishi[1,2]

[1]Department of Electronic Science and Engineering, Kyoto University, Kyoto, Kyoto 615-8510, Japan

[2]Center for Spintronics Research Network, Institute for Chemical Research, Kyoto University, Gokasho, Uji, Kyoto, 611-011, Japan

[3]Department of Materials Science and Engineering, Monash University, Clayton, Victoria 3800, Australia

[4]ARC Centre of Excellence in Future Low Energy Electronics Technologies, Clayton, Victoria 3800, Australia

[5]Robinson Research Institute, Victoria University of Wellington, Wellington 6140, New Zealand

[6]The MacDiarmid Institute for Advanced Materials and Nanotechnology, Wellington 6011, New Zealand

[7]PRESTO, Japan Science and Technology Agency, Honcho, Kawaguchi, Saitama 332-0012, Japan

**Corresponding authors**

[†]**Motomi Aoki**          E-mail: aoki.motomi.53r@st.kyoto-u.ac.jp

[†]**Masashi Shiraishi**     E-mail: shiraishi.masashi.4w@ kyoto-u.ac.jp

## ABSTRACT

Spin-orbit torque (SOT) is receiving tremendous attention from both fundamental and application-oriented aspects. $Co_2MnGa$, a Weyl ferromagnet that is in a class of topological quantum materials, possesses cubic-based high structural symmetry, the $L2_1$ crystal ordering, 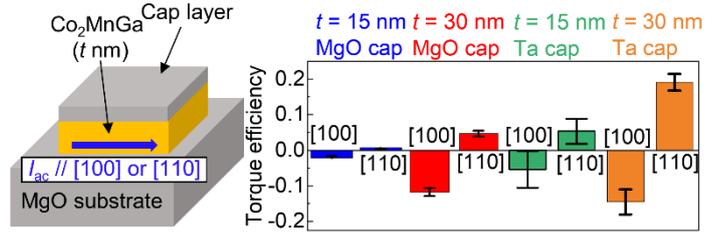 which should be incapable of hosting anisotropic SOT in conventional understanding. Here we show the discovery of a gigantic anisotropy of self-induced SOT in $Co_2MnGa$. The magnitude of the SOT is comparable to that of heavy metal/ferromagnet bilayer systems despite the high inversion symmetry of the $Co_2MnGa$ structure. More surprisingly, a sign inversion of the self-induced SOT is observed for different crystal axes. This finding stems from the interplay of the topological nature of the electronic states and their strong modulation by external strain. Our research enriches the understanding of the physics of self-induced SOT and demonstrates a versatile method for tuning SOT efficiencies in a wide range of materials for topological and spintronic devices.

**TEXT**

Spin-orbit interaction (SOI) has played a pivotal role for decades in both fundamental and application-oriented physics. Whereas the SOI generates a wide variety of intriguing physical effects, spin-orbit torque (SOT) in particular is one of the most important, attracting great attention for charge-spin conversion and manipulation of magnetization. Until recently, the magnitude of the SOI has been regarded as fixed for a given material, and uncontrollable. However, recent discoveries of tunable SOI in a nonmagnetic metal (Pt)[1] and an inorganic semiconductor (Si)[2] have countered this understanding and demand a revisit of SOI physics in view of controllability. Regarding the SOT, spin current generated by the spin Hall effect (SHE) in the bulk of nonmagnetic metals[3], due to interface SOI (ISOI)[4] and/or topological surface states[5] flows into an adjacent ferromagnetic material (FM), resulting in magnetization reversal. SOT has been demonstrated in a variety of systems, such as heavy metal/FM[6,7], topological insulator/FM[8,9], 2D material heterostructures[10], and even in single FM layers without adjacent materials[11,12], the latter referred to as self-induced SOT (SI-SOT). Despite the significant potential of SOT, its practical application in spintronic devices such as nonvolatile memory devices, oscillators, and domain-wall memories is still limited by the challenge of controlling its magnitude and orientation. Especially desirable for applications is a material with both large and inherently anisotropic SOT, because this would enable magnetization to be manipulated more flexibly in engineered structures that allow the electric current to be along different crystal axes. However, anisotropic SOT has so far been limited to low-symmetry materials[13–20]. Thus, to counter conventional understandings of SOT, one example that is worth investigating is the potential to manipulate SOT in highly symmetric materials, like cubic Heusler alloys, where substantial SOT has been observed due to the presence of Weyl nodes near the Fermi level[21–23].

Here, we report discovery of gigantic anisotropy in SOT in a single layer of Heusler alloy $Co_2MnGa$ (CMG), a Weyl ferromagnet[24,25], despite the hitherto understanding that its cubic-based $L2_1$ crystal ordering does not allow anisotropic SOT. The magnitude of the damping-like (DL) SOT efficiency, $\xi_{DL}$, as estimated using the second harmonic Hall (SHH) measurement, was comparable to that of heavy metal/FM bilayer structures, even without any adjacent metal layer. More surprisingly, the sign of $\xi_{DL}$ was reversed by switching electric-current orientation from the [100] to [110] crystal axes, indicating strong anisotropy of the SOT. Supported by theoretical calculations, additional experiments with different thicknesses of the CMG and different capping materials, have revealed an extraordinarily anisotropic DL SOT. This phenomenon, characterized by its remarkable anisotropy, stems from a complex interplay between the intrinsic topological properties of the electronic states and their modulation in response to the external strain induced by the substrate. Our research goes beyond conventional symmetry-anisotropy considerations for SOT and will have a great impact on SOT physics as well as novel device applications.

CMG (001) thin films with thicknesses of 15 and 30 nm were epitaxially grown on crystalline MgO (001) substrates with a MgO capping layer. Given that $a_{CMG} \sim \sqrt{2} a_{MgO}$, where $a_{CMG(MgO)}$ is the lattice constant of CMG (MgO), the cubic lattice of an epitaxially grown CMG film is rotated by 45° compared to the cubic MgO lattice as shown in Fig. 1(b). Figure 1(c) shows a $2\theta/\theta$ scan of X-ray diffraction (XRD) for a 30 nm CMG film. MgO 002, CMG 002, and CMG 004 peaks were obtained, indicating good crystalline order along the out-of-plane direction. In addition, we confirmed $L2_1$ ordering from the CMG 113 superstructure peak as shown in the inset of Fig. 1(c). Figures 1(d) and 1(e) show the result of $\varphi$ scans of XRD for the CMG 202 and MgO 202 peaks, respectively. Fourfold symmetry in 202 peaks and a 45° difference in $\varphi$ between CMG 202 and MgO 202 are consistent with the schematic shown in Fig. 1(b). Further growth and structural details can be found in our previous papers[21,26].

We used the SHH method to investigate the SOT in a CMG single layer[27,28]. Figure 2(a) shows a schematic of the measurement setup. The films were patterned into Hall bar structures and SHH voltage, $V_{2\omega}$, was measured using an AC current, $I_{ac}$ (see Supporting Information A for details). An external magnetic field, $B_{ext}$, was applied in the x-y plane at angle $\varphi$. The Hall bar structures were fabricated with their long directions along the [110] and [100] axes on the same substrate to investigate crystalline orientation dependence of the SOT. Figure 2(b) shows $V_{2\omega}$ as a function of $\varphi$ with $B_{ext}$ = 1 T for $t$ = 30 nm and $I_{ac}$ // [110]. The black line is fitted using,

$$V_{2\omega} = V_a \cos\varphi + V_{\text{FL}}(2\cos^3\varphi - \cos\varphi) + V_{\text{PNE}}\sin 2\varphi, \qquad (1)$$

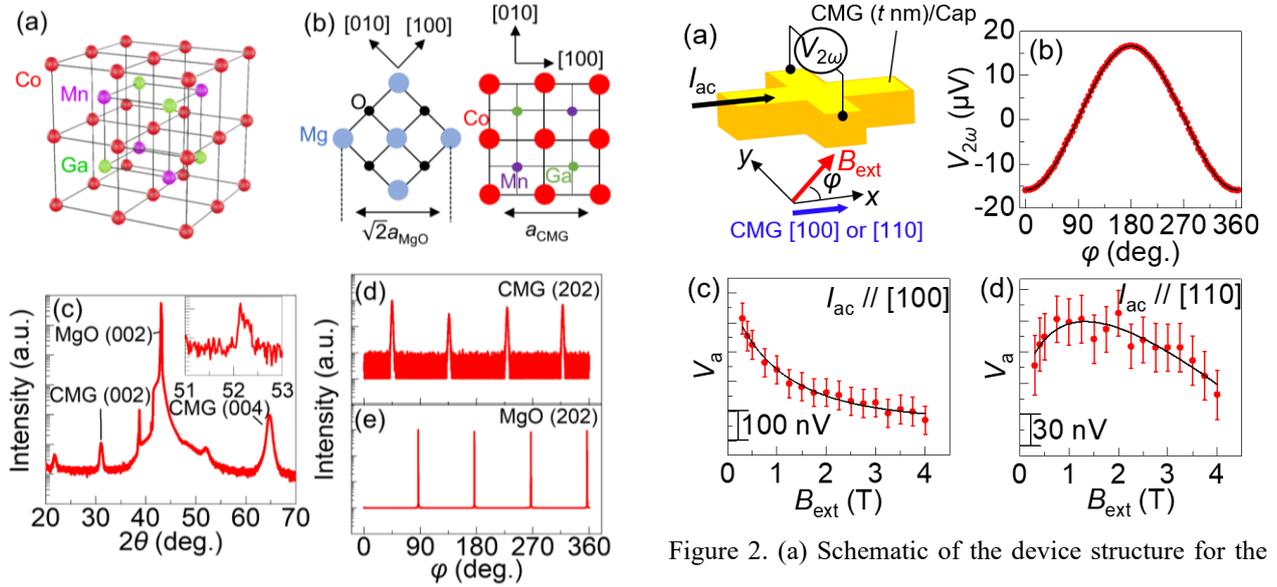

Figure 1. (a) Schematic of the crystal structure of L2$_1$ ordered CMG. (b) Schematic of the relationship between crystal orientation of CMG and MgO for our CMG film grown on MgO (001) substrate. (c) Spectrum of the $2\theta/\theta$ scan of the XRD. MgO 002, CMG 002, and CMG 004 peaks are observed, indicating single crystal growth along out-of-plane direction. Peaks at 22° and 39° originate from MgO substrate. Inset shows CMG 113 peak obtained with $\varphi$ = 45° and $\chi$ = 13.22°. (d) CMG 202 and (e) MgO 202 peaks obtained from the $\varphi$ scan of the XRD.

Figure 2. (a) Schematic of the device structure for the SHH measurement. (b) $V_{2\omega}$ as a function of $\varphi$ measured for the CMG(30 nm)/MgO(10 nm) device with $B_{ext}$ = 1 T. The black line indicates the fittings using Eq. (1). (c,d) $V_a$ as a function of $B_{ext}$ for the CMG(30 nm)/MgO(10 nm) device with (c) $I_{ac}$ // [100] and (d) $I_{ac}$ // [110]. Black lines are the fitting using Eq. (2). Error bars indicate standard error.

where $V_{FL}$ is the SHH voltage generated by the field-like SOT and the Oersted field, and $V_{PNE}$ is generated by the planar Nernst effect due to in-plane thermal gradient[27]. Here, $V_a$ is expressed as[28],

$$V_a = -\frac{V_{AHE} B_{DL}}{2(B_{ext} + B_k)} + A B_{ext} + V_{ANE}, \quad (2)$$

where $V_{AHE}$ is the voltage amplitude of the anomalous Hall effect (AHE), $B_{DL}$ is the effective field generated by the DL SOT, $B_k$ is out-of-plane anisotropy field, $A$ is the ordinary Nernst effect coefficient, and $V_{ANE}$ is the SHH voltage generated by the anomalous Nernst effect. Note that $B_{DL}$ is defined to be positive along $-z$ when an electric current is along $+x$ and $-90° < \varphi < 90°$ in our setup. To estimate field-like SOT we need to subtract the contribution from the Oersted field due to the nonuniformity of the electric current, which is difficult to calculate. Therefore, we only focus on the DL SOT in the following analysis. Figures 2(c) and 2(d) show $V_a$ as a function of $B_{ext}$ with $I_{ac}$ // [100] and [110], respectively. The black curves are the fitting done using Eq. (2). Here, $B_k$ was independently estimated from the measurement of the AHE. When $I_{ac}$ // [100], $V_a$ decreased as $B_{ext}$ was increased, whereas when $I_{ac}$ // [110] $V_a$ was enhanced up to 1 T. Given that the sign of the curvature of the fitting curve obtained by using Eq. (2) corresponds to the sign of $B_{DL}$, these results indicate a sign inversion of the DL SOT, as the sign of $V_{AHE}$ was always the same irrespective of the current orientation (see Supporting Information B).

The SHE is inactive in the non-magnetic and insulating MgO substrate and capping layer, so only possible source of spin current causing SOT is the CMG. Non-zero SOT is forbidden by symmetry for a centrosymmetric FM layer such as cubic Heusler alloy CMG, if there is no adjacent metal layer. To generate SI-SOT, inversion symmetry breaking along the out-of-plane direction is needed[11,29]. Here, we focus on the nonuniform distortion in the CMG film due to strain[29] imposed by the substrate. Because $\sqrt{2} a_{MgO}$ (5.961

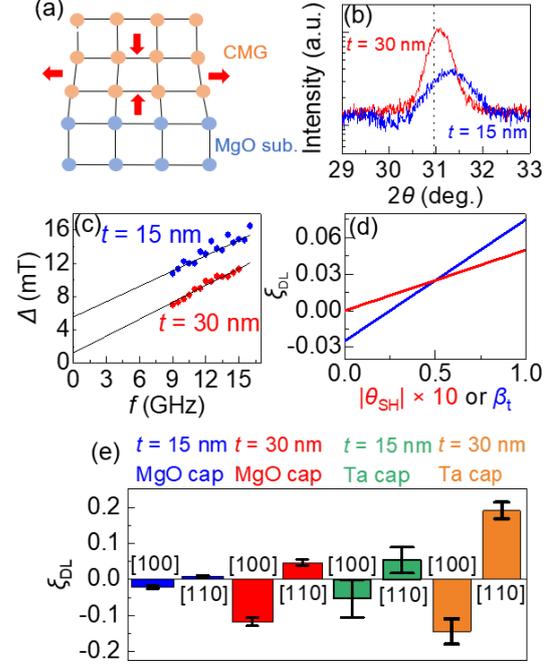

Figure 3. (a) A schematic of the nonuniform distortion due to the lattice mismatch between CMG and MgO. (b) CMG 200 peaks obtained from the $\theta$-$2\theta$ scan of the XRD for CMG/MgO films with $t$ = 15 nm and 30 nm. Out-of-plane lattice parameter for 15-nm and 30-nm CMG was estimated to be 5.753 Å and 5.711 Å, respectively. (c) $\Delta$ as a function of $f$ for CMG/MgO films with $t$ = 15 nm and 30 nm. (d) $\theta_{SH}$ and $\beta_t$ dependence of $\xi_{DL}$ calculated by using Eq. (3). $\beta_0 = 0.5$, $\theta_{SH(t)} = 0.1$, and $\theta_{SH(0)} = 0.05$ in the calculation of $\beta_t$ dependence, whereas $\beta_0$ and $\beta_t$ are set to be 0.5 in the calculation of $|\theta_{SH}|$ dependence. (e) $\xi_{DL}$ for CMG (15 nm)/MgO (10 nm), CMG (30 nm)/MgO (10 nm), CMG (15 nm)/Ta (3 nm), and CMG (30 nm)/Ta (3 nm) measured by the SHH method. Error bars indicate standard error.

Å) is slightly longer than the bulk value of $a_{CMG}$ (5.767 Å)[30], a tetragonal distortion is expected as shown in Fig. 3(a)[31]. The distortion in the CMG relaxes away from the interface with the MgO substrate, resulting in a thickness dependent spin Hall angle (SHA), allowing for SI-SOT. Here, the spin current generated by the AHE[32], which is generated due to the SOI and spin-polarized current in the FM, cannot give rise to DL SOT because the spin polarization, $\sigma$, is parallel to the magnetization, $M$[33]. Therefore, we only focus on the magnetization independent SHE[33–35], where $\sigma$ deviates from $M$. In this case, the effective spin current, $J_s$, at the substrate/CMG and CMG/cap interfaces is expressed as $J_s(t) = J_c\theta_{SH}(0)\beta_0$ and $J_s(t) = J_c\theta_{SH}(t)\beta_t$, respectively, where $\theta_{SH}(0)[(t)]$ is the SHA of CMG in the region near the substrate (cap) and $\beta_{0(t)}$ is the spin absorption rate into the substrate/CMG (CMG/cap) interface (see Supporting Information C). Note that $t \gg \lambda$ is assumed in the above expression, where $\lambda$ is the spin relaxation length (~ 3 nm in CMG[21]). Non-zero DL torque efficiency, $\xi_{DL}$, is generated[36] because

$$\xi_{DL} \sim \frac{\int_0^t \nabla J_s(x)dx}{J_c} = \frac{J_s(t) - J_s(0)}{J_c} = \theta_{SH}(t)\beta_t - \theta_{SH}(0)\beta_0 \neq 0. \quad (3)$$

To confirm the existence of the strain, first, we compared XRD spectra for different film thicknesses. Given that the lattice mismatch between the CMG film and the MgO substrate will cause a strain, thinner CMG is expected to show a stronger distortion. Figure 3(b) shows the CMG 002 peaks obtained from the $\theta$-$2\theta$ scan of XRD for different thickness films. A dotted line indicates the expected peak position assuming no distortion to the bulk CMG lattice. For $t$ = 30 nm the measured CMG 002 peak is at a higher angle and it shifts even higher angle at lower $t$, confirming the out-of-plane lattice parameter is smaller than for bulk CMG. Assuming the volume of the unit cell is not changed by distortion, the in-plane lattice must be expanded and the average lattice constant along the in-plane direction is ~ 5.774 Å and 5.795 Å for $t$ = 30 nm and 15 nm films, respectively. Both values are between $\sqrt{2}a_{MgO}$ and $a_{CMG}$, consistent with the expectation in Fig. 3(a). Figure 3(c) shows the line width of the ferromagnetic resonance (FMR), $\Delta$, as a function of microwave frequency, $f$, obtained from the spin-torque FMR method[37] and the black lines are the linear fitting. It is known that the inhomogeneous damping, i.e., the intercept of the linear fit with the $y$ axis, is enhanced by lattice distortion[30]. The higher value of the inhomogeneous damping for the thinner CMG film means it is more distorted compared with the thicker one, which agrees with the model[30,31].

If the origin of the SOT is non-uniform distortion in the CMG, causing a thickness dependent SHE, the SOT will also be CMG-thickness and capping-layer dependent because they influence $\theta_{SH}(t)$ and $\beta_t$, respectively. To confirm this, we investigated devices with a different $t$ and a different capping layer. Figure 3(d) shows the evolution of $\xi_{DL}$ as a function of $|\theta_{SH}| = \theta_{SH}(t) - \theta_{SH}(0)$ and $\beta_t$ calculated using Eq. (3). Here, we assumed $\theta_{SH}(t) > \theta_{SH}(0)$. Given that $\xi_{DL}$ monotonically increases with both parameters, in both devices with a thicker CMG film, which results in an increment of $|\theta_{SH}|$, and devices with a capping material with larger spin absorption rate, e.g., a heavy metal, $\xi_{DL}$ is expected to increase. Figure 3(e) shows $\xi_{DL}$ for CMG (15 nm)/MgO (10 nm), CMG (30 nm)/MgO (10 nm), CMG (15 nm)/Ta (3 nm), and CMG (30 nm)/Ta (3 nm) measured by the SHH method. We used the relationship between $B_{DL}$ and $\xi_{DL}$[38],

$$B_{DL} = \xi_{DL} \frac{\hbar I_{ac}}{2eM_s w t^2}, \qquad (4)$$

where $M_s$ and $w$ are saturation magnetization obtained by a vibration sample magnetometer (see Supporting Information D), and width of the Hall bar, respectively. $\xi_{DL}$ was enhanced in both the thicker device and the Ta-capped device, which agrees with the above expectations.

To further support the hypothesis of a strain-dependent SHE in CMG and to confirm that this causes the anisotropic SI-SOT, we carried out first principles calculations (see Supporting Information A for details [39–43]). Figures 4(a)-(d) show spin Hall conductivity (SHC), $\sigma_s$, calculated for CMG with and without tetragonal distortion, plotted as a function of $E - E_F$ where $E$ and $E_F$ are the electronic state energy and the Fermi level, respectively. We integrated the Berry curvature (BC) only along certain directions to clarify the electric-current direction dependence of $\sigma_s$. Along the [100] direction, whereas $\sigma_s$ of the unstrained CMG is dominated by contributions via the spin-down channel [Fig. 4(a)], $\sigma_s$ of the strained CMG is mostly dominated by spin-up channel contributions [Fig. 4(b)], indicating that $\sigma_s$ is sensitive to the lattice distortion. Along the [110] direction, $\sigma_s$ is also strongly modulated by lattice distortion. The spin-up channel dominates near the Fermi level when the lattice is unstrained [Fig. 4(c)], while the strained CMG leads to $\sigma_s$ peaks mainly in the spin-down channel [Fig. 4(d)]. Therefore, $\sigma_s$ depends heavily on the crystal direction used for integration of BC in both strained and unstrained CMG. When we focus on $\sigma_s$ within $|E - E_F| < kT \sim 25$ meV, where $T$ is temperature, $\sigma_s$ of strained CMG is much larger than that of unstrained CMG when BC is integrated along the [100] axis, while the relationship is inverted when BC is integrated along the [110] axis. Considering that DL SOT is given by Eq. (3), these calculations well reproduce the sign inversion of the DL SOT measured along the different axes as $\theta_{SH}(t) < \theta_{SH}(0)$ ($\theta_{SH}(t) > \theta_{SH}(0)$) when BC is integrated along [100] ([110]). We note that this result also explains the fact that $|\xi_{DL}|$ of Ta-capped CMG is enhanced, especially when $I_{ac}$ was applied along [110] [see Fig. 3(e)]. In case of $I_{ac}$ // [100], $|\xi_{DL}| \sim |\theta_{SH}(0)\beta_0| - |\theta_{SH}(t)\beta_t|$, resulting in smaller $|\xi_{DL}|$. On the other hand, in case of $I_{ac}$ // [110], $|\xi_{DL}| \sim |\theta_{SH}(t)\beta_t| - |\theta_{SH}(0)\beta_0|$ [see Eq. (3)], resulting in enhancement of $|\xi_{DL}|$ owing to larger $\beta_t$.

Using the band structure calculation, we can understand how the modulation of $\sigma_s$ by tetragonal distortion and the strong integration-axis dependence of $\sigma_s$ both rely on the topological nature of the electronic states in CMG. The large $|\sigma_s|$ of CMG is due to BC hot spots corresponding to the Weyl points in the band structure[24–26]. Since BC hot spots concentrate along specific crystal directions (see Supporting Information E), $\sigma_s$ also strongly depends on the crystal orientation. Second, the position of the Weyl points is significantly changed by the tetragonal distortion. Therefore, the distortion also modifies the corresponding BC hot spots and $\sigma_s$ (see Supporting Information E). This contrasts FM without topological electric states, such as Fe, which has no Weyl points near the Fermi level. Hence, although strain would be expected in a thin layer of Fe, only a small DL SOT and an insignificant SOT anisotropy is measured (see Supporting Information F). From the above experimental and theoretical results, we conclude that large SI-SOT in CMG is induced by the lattice mismatch between CMG and MgO, resulting in a tetragonal distortion causing strong modulation of the Weyl points, and the DL SOT strongly depends on the electric-current direction as Weyl points are dense along specific crystal axes.

SHE in a high-symmetric material is usually independent of the crystal-axis direction, i.e., $\sigma_s$ is the same irrespective of the applied electric-current direction[13]. This accounts for the fact that anisotropy in the SHC, $|\Delta\sigma_s|$, of fcc Pt is small[20], whereas those of lower symmetry structures like hexagonal $Mn_3Ga$[16] and orthorhombic $SrIrO_3$[44] are large, as shown in Table 1. Despite the high-symmetry crystal structure, $|\Delta\sigma\xi_{DL}| = 1.4 \times 10^5$ $(\Omega m)^{-1}$ is achieved in CMG, much larger than $|\Delta\sigma_s|$ in fcc Pt and even larger than low-symmetry materials as shown in Table 1. Our results indicate that large anisotropy can be achieved even in a high-symmetry material if there are Berry curvature hot spots along specific symmetry lines that can be modulated by external strain. Strain caused by lattice mismatch exists in almost all thin films unless they are grown on perfectly lattice matched substrates, making our findings widely applicable and could be a mechanism for introducing SOT anisotropy in many other materials, including other high-symmetry ones.

Table 1. Summary of the anisotropy of SOT.

|  | Pt[20] | $Mn_3Ga$[16] | $SrIrO_3$[44] | CMG (This work) |
|---|---|---|---|---|
| Crystal structure | fcc | Hexagonal | Orthorhombic | $L2_1$ |
| $|\Delta\sigma\xi_{DL}|$ ($\times 10^4$ $\Omega^{-1} m^{-1}$) | 1.5 | < 6.0 | 4.0 | 14 |

Given that $|\sigma_s|$ gives upper limit of $|\sigma\xi_{DL}|$, we used $|\Delta\sigma_s|$ as an upper limit of $|\Delta\sigma\xi_{DL}|$ for $Mn_3Ga$. $|\Delta\sigma_s|$ is taken as the difference between the SHC with $I_{ac}$ // [100] and that with $I_{ac}$ // [110].

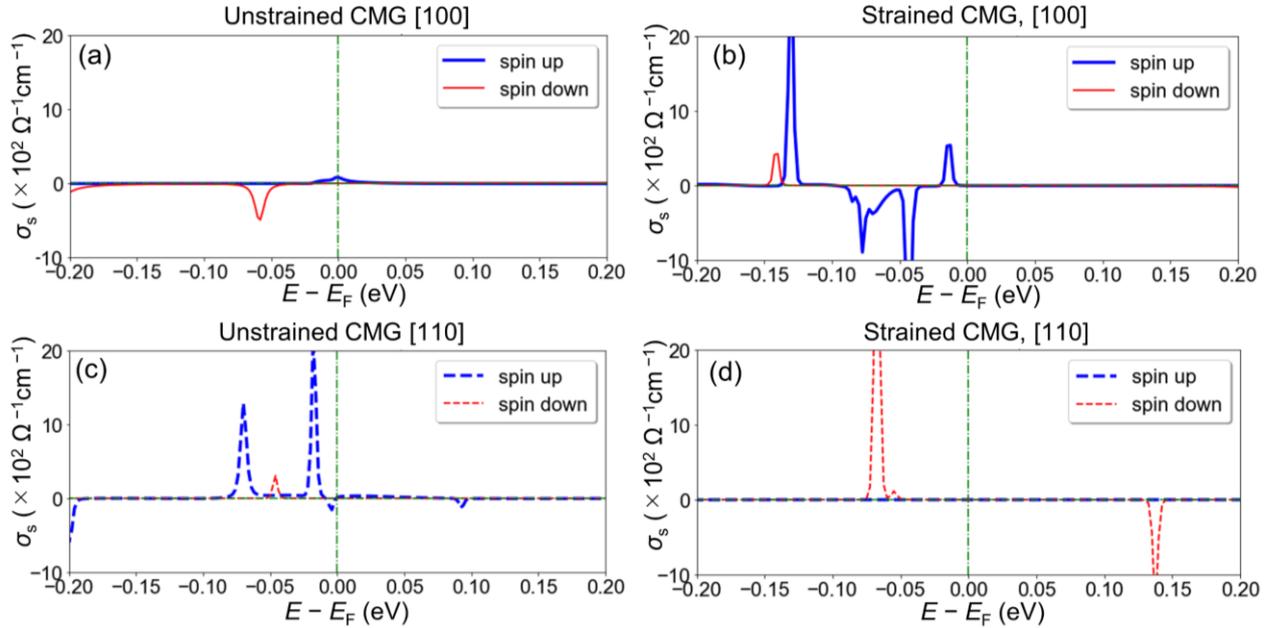

Figure 4. $\sigma_s$ as a function of $E - E_F$ for (a) unstrained and (b) strained CMG when BC is integrated along [100], and (c) unstrained and (d) strained CMG when BC is integrated along [110].

Next, we comment on the sign and magnitude of the SHA in CMG estimated from our experimental results. In our measurement setup, positive $\xi_{DL}$ corresponds to positive (negative) spin-current injection from the bottom (top) interface. Given that for the [100] ([110]) direction $\theta_{SH}(t) < \theta_{SH}(0)$ ($\theta_{SH}(t) > \theta_{SH}(0)$), sign of $\xi_{DL}$ corresponds to that of $\theta_{SH}(0)\beta_0$ ($-\theta_{SH}(t)\beta_t$), our experimental anisotropic SOT result is the consequence of negative SHA in both cases [see Fig. 3(e)]. We cannot estimate the precise value of $\theta_{SH}$ because of the unknown $\beta$ in Eq. (3), however, $\theta_{SH} < -0.15$ along [100] and $\theta_{SH} < -0.19$ along [110] is expected from the values of $\xi_{DL}$ in the Ta-capped device in Fig. 3(e). We previously reported $\theta_{SAH} + \theta_{SH} = -0.19$ using a spin valve structure and transport measurement with electric current along the [110] axis[21], where $\theta_{SAH}$ is the spin anomalous Hall angle[32,33]. Isshiki et al. reported $\theta_{SAH} + \theta_{SH}$ to be $-0.09$ with careful subtraction of thermoelectric artifacts[22]. Tang et al. reported that the contribution from $\theta_{SH}$ is large compared with $\theta_{SAH}$ from the SHH measurement in CMG/Ti/CoFeB layers[45]. These facts indicate that the $\theta_{SH}$ in CMG is in the order of $-0.1$, consistent with our measurement.

We also need to comment on the interface spin absorption rate, $\beta$, because it is essential for generating SI-SOT [see Eq. (3)]. Generally, $\beta$ is strongly correlated with interface magnetic anisotropy because both come from the ISOI[46]. Especially, transition metal/oxide interfaces such as MgO/CMG have strong ISOI owing to the bonding between oxygen and metal ions[47,48]. In MgO/CMG/Pd layers with perpendicular magnetic anisotropy, both MgO/CMG and CMG/Pd interfaces take a crucial role, indicating that the MgO/CMG interface has a strong interface magnetic anisotropy energy[49]. These facts well explain the sizable $\beta$ at the MgO/CMG interface and resulting SI-SOT.

Finally, we briefly discuss other possible origins for the DL SOT in our CMG films. One might expect charge-to-spin conversion at the top and/or bottom interface via Rashba SOI causing DL SOT[47]. However, this cannot explain the enhancement of $\xi_{DL}$ as increasing $t$ for the CMG films capped with Ta, because asymmetry in the Rashba SOI between MgO/CMG and CMG/Ta interfaces is independent of $t$. Therefore, DL SOT in our CMG is a bulk phenomenon. Although AHE itself cannot give rise to SI-SOT, a combination of AHE and spin rotation via surface scattering can[50]. However, this kind of SOT is scaled by $\lambda$ (~3 nm in CMG), which cannot explain the strong enhancement of $\xi_{DL}$ in the thicker CMG layer. Similarly, anomalous SOT is not the origin of our results[11].

In conclusion, we found that the sign of the DL SOT in single CMG layers is inverted by switching the electric-current direction from the [100] to [110] crystal axis, the origin of which is the interplay between the topology of the electronic states and the strain exerted on the films by the MgO substrate with non-identical lattice parameter. The anisotropy of the DL SOT exceeds any high- and low-symmetry materials reported so far. Our finding overcomes the conventional understanding that anisotropy in SOT is necessarily small in high-symmetry materials and provides a method for energy efficient SOT devices to be realized by designing structures with proper external stain and optimized electric current orientation.

**ASSOCIATED CONTENT**

**Supporting Information**

The Supporting Information is available free of charge.

Fabrication, measurement, calculation details, anomalous Hall effect measurement, model for spin-orbit torque in a ferromagnetic single layer, saturation magnetization measurement, band structure calculation, and control experiment using Fe.

**AUTHOR INFORMATION**

**Author contributions**

M.A. and Y.A. conceived the original concept. M.A. fabricated devices and collected data with help from L.L. and R.O. S.G. and Y.Z. grew films. Y.Y. carried out theoretical calculations under the supervision of N.V.M. M.A., S.G., Y.Y., N.V.M, and M.S. analyzed the results. M.A. wrote the manuscript with help all the authors. All the authors discussed the results. S.G. and M.S. guided the project.

**ACKNOWLEDGEMENTS**

M. A. acknowledges support from JSPS Research Fellow (Grant No. 22J21776). S.G. acknowledges financial support from the New Zealand Science for Technological Innovation National Science Challenge. The MacDiarmid Institute is supported under the New Zealand Centres of Research Excellence Programme. Y.Y. and N.V.M. acknowledge the support from the ARC Centre of Excellence in Future Low-Energy Electronics Technologies (Grant No. CE170100039), and the computational support from the National Computing Infrastructure, Australia, and the Pawsey Supercomputing Centre.